\documentclass{article}
\usepackage{authblk}
\usepackage[utf8]{inputenc}
\usepackage{graphicx}
\usepackage{pdfpages}
\usepackage{dcolumn}
\usepackage{bm}
\usepackage{amsmath}
\usepackage{amssymb}
\usepackage{array}
\usepackage{multirow}
\usepackage{color}
\usepackage{sidecap}
\usepackage{float}
\usepackage{hyperref}
\usepackage{setspace}
\usepackage{geometry}
\usepackage[backend=biber, style=chem-acs, citestyle=chem-acs, doi=true, articletitle=true, chaptertitle=true, url=false,isbn=false, arxiv=true]{biblatex} 
\AtEveryBibitem{
  \clearfield{doi}
  \clearfield{url}
  \clearfield{isbn}
}
\usepackage{mathspec}

\title{\bf Polymer Knots in Thin Films: Thickness Dependence, Local Effects, and Stiffness}

\author[1]{M. P. Schmitt$^*$}
\author[2]{H. Meyer}
\author[1]{P. Virnau}
\affil[1]{Institut für Physik, Johannes Gutenberg-Universität Mainz, Staudingerweg 9, 55099 Mainz, Germany}
\affil[2]{Institut Charles Sadron, Université de Strasbourg, CNRS UPR 22, 23 rue du Loess-BP 84047, 67034 Strasbourg, France}

\begin{document}


\maketitle

\begin{abstract}
    We study how confinement affects topology and conformations in polymer films of varying thickness $h$. The knotting probability exhibits a maximum at intermediate thicknesses near the bulk radius of gyration $h \approx R_\mathrm{g,bulk}$, vanishes at small $h$ and approaches bulk values for large $h$. Close to walls, the entanglement length increases monotonically and conformations become flatter. A layer-resolved analysis of structural and topological properties allows us to reconstruct the explicit thickness dependencies by integrating layer-resolved properties of a thick film.
\end{abstract}

\section*{Introduction}
Confinement is ubiquitous in both natural and technological settings: DNA and proteins are compressed within cellular compartments or viral capsids \cite{Micheletti2006-mg, dai:Macromol:2018,  zhou2008protein, reith2012effective, Molecular_Biology_of_the_Cell}, 
while synthetic polymers are often processed in thin films for coatings, membranes, and electronic applications \cite{kausar2018polymer, ulbricht2006advanced, mai2015ferroelectric, katz2009thin}. 
A key question is how reduced dimensionality and the presence of boundaries modify the spectrum of accessible chain conformations, and in particular, the occurrence of knots and entanglements \cite{Dai:Macromol:2014, ruskova2025controlling}.
{Although these two terms are sometimes used interchangeably, we underline that knots and entanglements are two different concepts. We consider a knot in the mathematical sense as a property of a single closed loop. Applied to linear polymers, an appropriate closure is required. An entanglement occurs when two chain segments close in space hinder the perpendicular motion of each other. Entanglements (in the rheological sense) occur as soon as many long enough chains share the same volume; they are a multi-chain property.} \newline

In polymer science, interest in knots dates back to the Frisch–Wasser\-man–Delbrück conjecture \cite{Frisch:JACS:1961, Delbruck_knot_62} of the early 1960s, which asserts that sufficiently long polymer chains are generally knotted, consistent with everyday observations of macroscopic ropes and strings. Over the past five decades, numerical simulations—primarily of single polymer chains—have revealed, however, that the length scale at which knotting typically occurs, is highly sensitive to microscopic chain properties \cite{Vologodskii:1974, frank1975statistical, koniaris_1990, vanrensburg1992entanglement, Mansfield:Macro:1994, Deguchi1997, PhysRevE.61.5545, Grosberg:PRL:2000, orlandini2003polymer, marcone2005length, Kantor, Micheletti2006-mg, sulkowska2009dodging, virnau2010, Micheletti_PhRep2011, reith2012effective, Virnau:BCST:2013, Trefz:PNAS:2014, Wuest:PRL:2015, Dai:Macromol:2014, Dai:PRL:2015, rieger2016monte, najafi2016role, Marenz_PRL2016, Shimamura:JPA:2000, Rybenkov:1993:PNAS}. Short, flexible ideal chains readily form numerous \cite{Vologodskii:1974, Deguchi1997, Grosberg:PRL:2000}, strongly localized knots \cite{PhysRevE.61.5545}. Incorporating excluded-volume interactions shifts the onset of knotting to chain lengths larger by roughly two orders of magnitude, as quantified, for example, by a fixed knotting probability \cite{frank1975statistical, koniaris_1990, Deguchi1997, Kantor}. Although these knots remain weakly localized \cite{orlandini2003polymer, marcone2005length, Kantor}, their spatial extent is substantially larger than in the ideal case, underscoring the nontrivial role of microscopic interactions. This complexity is further highlighted by recent studies \cite{Coronel_2017, Uehara_2019} showing that the knotting probability of self-avoiding ring polymers varies non-monotonically with persistence length when contour length and excluded-volume size are held fixed. Similar non-monotonic behavior has also been reported for single rings \cite{tesi1994knot}, circular DNA \cite{dai2012effect, Micheletti:Macromol:2012} 
and linear DNA \cite{orlandini2013knotting, Wettermann_2023} with respect to planar confinement. In contrast, globular polymers and polymers confined within spherical cavities are typically highly knotted \cite{Mansfield:Macro:1994, Kantor}, but their knots are loose and delocalized \cite{orlandini2003polymer, Kantor}, effectively becoming global features of the chain. \newline

By comparison, knotting in multichain systems has been explored only sparsely \cite{Foteinopoulou:PRL:2008, Laso:SM:2009, trefz2015scaling, Meyer2018, Tubiana_2021, Jianrui2020, ubertini2023spatial, schmitt2024topological, Knots_in_soft_matter_review}. For polymer melts, one might invoke the Flory hypothesis \cite{Flory1949}, which predicts that excluded volume interactions are screened beyond the Edwards correlation length, yielding ideal chain statistics at large scales. However, recent simulations \cite{wittmer2007b, wittmer2007, glaser2014collective} demonstrate that this picture breaks down at the topological level. Knots in polymer melts occur significantly less frequently and are looser than in equivalent isolated ideal chains, an effect most pronounced for flexible polymers where local structure is crucial \cite{Jianrui2020}. Notably, polymers at the $\Theta$-point exhibit similar behavior, indicating that even when global conformational statistics appear ideal, topological properties can deviate markedly from random walk expectations \cite{schmitt2024topological}.
\newline

{Contrary to knotting, chain conformations and entanglements in thin polymer films have been subject to many studies \cite{brown1996entanglements,CaMuWiJoBi2005jpcm,MeKrCaWiBa2007epjst,SussmanEtal2014macro,lee2017local,PrRiWi2019macro}. The Silberberg argument supposes that a random walk hitting a hard surface is just reflected. From this, one can conclude that perpendicular dimensions are reduced and in-plane dimensions remain unchanged. Brown and Russel \cite{brown1996entanglements} argued that by the reflection the self-density of a chain close to a wall is increased, leaving less space for other chains in the pervaded volume, and consequently reducing the entanglement density. This perturbation has been quantified and tested in Ref. \cite{SussmanEtal2014macro}.
Semenov and Johner \cite{SeJo2003epje} pointed out that for strong confinement (film thickness smaller than the radius of gyration) with excluded volume, there must also be corrections to chain ideality and the in-plane dimensions will swell logarithmically \cite{CaMuWiJoBi2005jpcm,MeKrCaWiBa2007epjst}. This has been used in Ref. \cite{lee2017local} to calculate the entanglement length for polymer melts in slit confinement. The current work allows to test their prediction on the standard Kremer-Grest model.}
\newline

The interplay between knotting and slit-pore confinement in dense polymer films remains largely unexplored: 
To our knowledge, only one study examines semiflexible closed rings on a lattice under such conditions \cite{ubertini2023spatial} and finds that the knotting probability increases when the film thickness is comparable to the bulk size of the rings (when rings are allowed to change topology via Monte Carlo moves). Even less is known about how local, layer-resolved topological and structural properties inside a film correlate with global behavior. Since confinement introduces heterogeneity across the film profile, one may ask whether the topological statistics of thin films can be predicted from local chain properties in thicker systems. The role of stiffness in knot formation in slit-pore confinement has also not been addressed. \newline

In this work, we address these questions through extensive molecular dynamics simulations of polymer films with varying thickness. We analyze both entanglements and knots, focusing on their dependence on the confining dimension $h$ relative to the bulk radius of gyration $R_\mathrm{g,bulk}$. Similar to observations for single chains, we find that the knotting probability is non-monotonic, exhibiting a maximum around $h \approx R_\mathrm{g,bulk}$ and vanishing in the $2d$ limit, whereas entanglements decrease monotonically with confinement. Layer-resolved analysis in thick films reveals that the {modulation} of $R_\mathrm{g}$ and enhanced knotting near the walls mirror the global dependence on $h$, suggesting that thickness-dependent properties can be reconstructed from local information. Comparisons with single confined chain Monte Carlo simulations highlight the role of screening by neighboring chains in dense films, and simulations with semiflexible polymers suggest that stiffness shifts the knotting maximum to smaller $h$, providing a mechanistic explanation for its absence in our accessible range. \newline 

In summary, these results establish a coherent framework for understanding how confinement, topology, and chain flexibility interact in polymer films. They highlight the bulk radius of gyration as the key length scale governing both local and global confinement effects, and they provide new methodological and conceptual tools for predicting topological behavior in thin polymer films.

\section*{Methods}
\subsection*{Polymer Simulations}
Polymer film configurations are generated using molecular dynamics (MD) simulations based on the LAMMPS \cite{PLIMPTON19951} software package. Bonds are modeled by a harmonic potential with total energy $V_\text{harm} = \sum_{i=1}^{N-1} A_\mathrm{harm} \cdot (r_{i,i+1} - l_b)^2$ for each chain with $N$ beads, where $r_{i,i+1}$ is the distance between neighboring monomers $i$ and $i+1$ along the chain. The expectation value of the bond length 
is $l_b = 0.967 \, \sigma$, while $A_\mathrm{harm} = 400 \, k_B T/\sigma^2$ provides a large energy penalty for deviations from $l_b$. 
The qualitative behavior of the resulting bonds is in line with the standard FENE potential \cite{KremerGrest}. $k_B T = 1$ is the energy scale of the system and $\sigma = 1$ is the unit of length in simulations. 
Non-neighboring monomers experience a purely repulsive Weeks-Chandler-Andersen (WCA) potential \cite{WCA}. The WCA potential is a Lennard-Jones potential, which is cut off at the minimum, such that the energy contribution per non-neighboring pair is: $V_\text{WCA} = 4\epsilon \left( \frac{\sigma^{12}}{r^{12}} - \frac{\sigma^6}{r^6} \right) + c$ if $r < r_c = \sqrt[6]{2} \sigma$, and $0$ otherwise. The constant $c =  -4\epsilon \left( \frac{\sigma^{12}}{r_c^{12}} - \frac{\sigma^6}{r_c^6} \right)$ enforces that $V_\text{WCA}$ is a continuous function, and is $\epsilon = 1 \, k_B T$ for the WCA cutoff $r_c = \sqrt[6]{2} \, \sigma$. This interaction applies to non-neighboring monomer pairs of the same chain (intra-chain pair interaction) as well as monomers of different chains (inter-chain pair interaction). 
The flexibility of the chain is modeled via a Kratky-Porod type potential in line with Ref. \cite{schmitt2024topological}. The potential depends on the angles formed by two consecutive bonds $\theta_i = \arccos \left( \hat r_{i-1,i} \cdot \hat r_{i,i+1} \right)$, where $\hat r_{i,i+1}$ is the unit vector pointing 
from monomer $i$ to monomer $i+1$. The total bending energy of a chain is given by $V_{B} = \sum_{i=2}^{N-1} B\cdot(1-\cos \theta_i)$ and the prefactor $B$ controls the stiffness of the chain. It is varied from the flexible case $B=0$ up to the semiflexible case $B=2$. 
Film confinement is enforced by using two flat walls at a distance $h$ apart in the $y$-direction, with all monomers positioned between the walls. The repulsive potential of the walls is the same as the purely repulsive pair-potential, i.e., it is $V_\text{wall} = 4\epsilon \left( \frac{\sigma^{12}}{\Delta y^{12}} - \frac{\sigma^6}{\Delta y^6} \right) + c$ if $\Delta y < r_c = \sqrt[6]{2} \sigma$, where $\Delta y$ is the absolute $y$-distance of the monomer to the considered wall. 
The system is evolved over time with the standard Langevin integrator of LAMMPS, which includes random forces in line with a $NVT$-ensemble. \newline  

For comparison we also perform fixed bond length ($l_b = 0.967$) reptation Monte Carlo simulations \cite{binder1995monte}. To emulate the behavior of $\theta$-chains we allow for attractive Lennard-Jones interactions with a cutoff of $r_c = 2 \cdot \sqrt[6]{2}$, but set $\epsilon_\mathrm{LJ}$ to $0.334$ in line with Ref. \cite{schmitt2024topological}.

\subsection*{Knot and Entanglement Analysis}
We use a custom code to detect knot types and lengths, as outlined in Ref. \cite{virnau_2010}. Strictly speaking, knots are only mathematically defined in closed loops. Therefore, a closure is required to artificially close our open chains. This is done by extending the chain ends far outwards from the center of mass of the polymer before connecting them. The code then reduces the structure while keeping topological constraints intact using the KMT reduction scheme \cite{koniaris_1990, koniaris_1991, Taylor_2000}. 
Finally, the Alexander polynomial \cite{alexander1928topological} of a planar projection of the trajectory is computed, which suffices to identify simple knots. The length of simple non-trivial knots is computed by iteratively removing beads from the chain ends until the knot type changes. Knotting probabilities and layer-resolved analyses are finally obtained by counting the occurrences of non-trivial knots and binning data with respect to their $y$-position, respectively. \newline

To characterize entanglements within polymer films, we employ the Z1+ code \cite{kroger2023z1+}. The method identifies and minimizes the primitive paths of individual polymer chains under topological constraints, thereby quantifying entanglement points without altering the underlying molecular connectivity. 
The resulting entanglement lengths $N_\mathrm{entang}$ are used to assess the influence of confinement on chain topology. Specifically, we use the modified kink-based entanglement length output by the Z1+ code, which is further explained in Ref. \cite{kroger2023z1+}.

\section*{Results and Discussion}

\subsection*{Film-Thickness Dependence of Polymer Topology: Knotting Probability, Entanglement Length, and Trefoil Size}

\begin{figure}[h!]
    \centering
    \includegraphics[width=.475\textwidth]{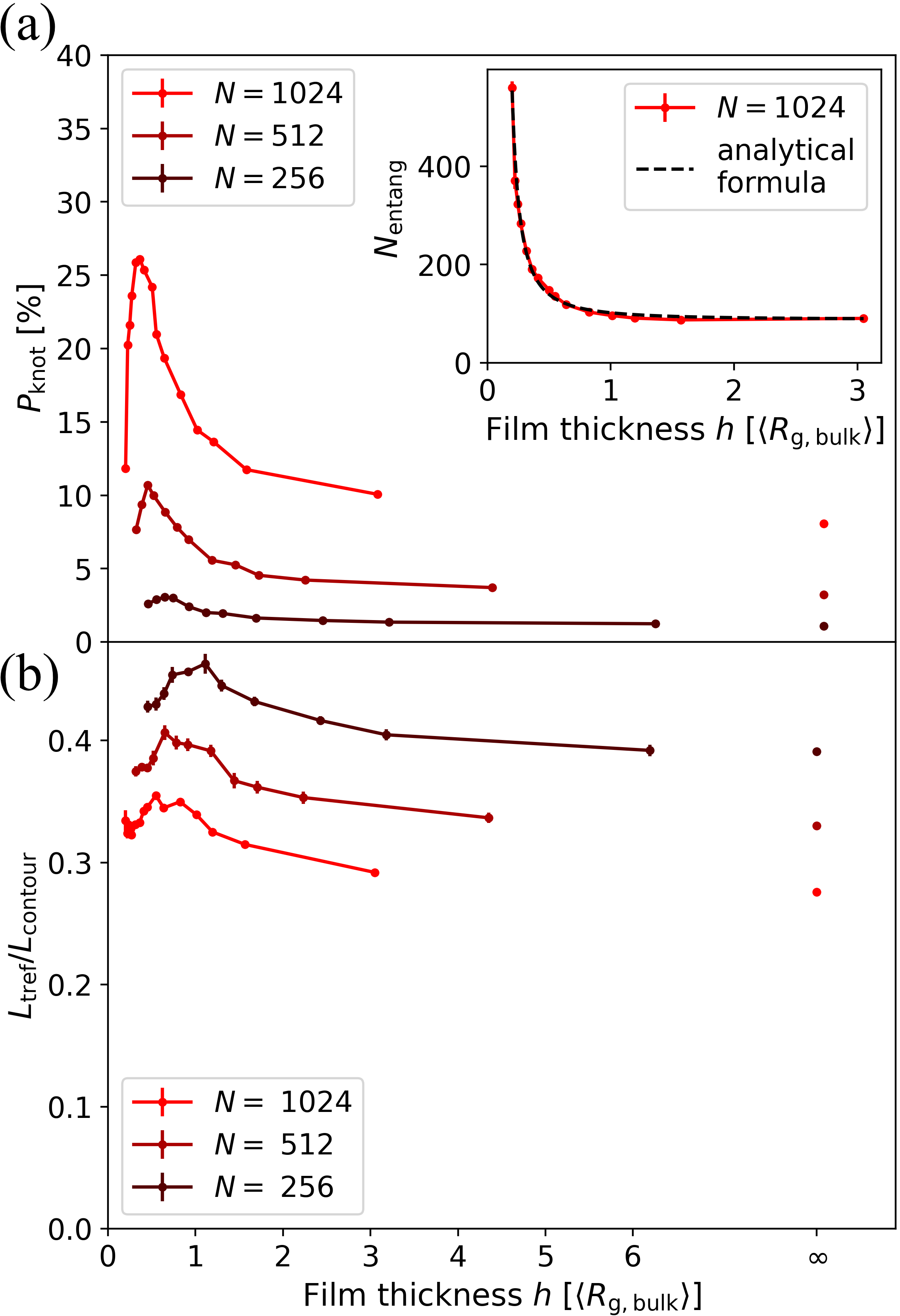}
    \caption{\textbf{Knotting probability $P_\mathrm{knot}$, entanglement length $N_\mathrm{entang}$ and normalized trefoil knot lengths $L_\mathrm{tref}/L_\mathrm{contour}$ of confined films as a function of the film thickness $h$.} {Thicknesses are normalized by the polymers' bulk radii of gyration $R_\mathrm{g,bulk}$ and} all points correspond to a {bulk} monomer density of $\rho = 0.68 \, \sigma^{-3}$. (a) The knotting probability exhibits a maximum at a film thickness of $h \approx 0.6 \cdot R_\mathrm{g,bulk}$ for all chain lengths. The knotting probabilities of bulk melts for each chain length are indicated by $h = \infty$; the knotting probabilities of the films consistently approach these bulk values. In contrast, the limit $h \rightarrow 0$ leads to decreasing knotting probabilities. The inset shows the entanglement length, which monotonically decreases with increasing thickness, along with a fit to an analytical expression from Ref. \cite{lee2017local}. (b) The trefoil knot lengths exhibit maxima at $h \approx  R_\mathrm{g,bulk}$. }
    \label{Pknotvsfilmthickness_films}
\end{figure}

Fig. \ref{Pknotvsfilmthickness_films}(a) shows the non-monotonic knotting probability as a function of the film thickness $h$. Introducing confinement down to $h \approx R_\mathrm{g,bulk}$ increases the knotting probability since internal crossings are more likely to occur as chains become more compact. The sharp decrease in knotting probability in the case of stronger confinement $h \ll R_\mathrm{g,bulk}$ can be understood by considering that the chains approach the $2d$-limit, where chains can no longer form crossings. Therefore, knot formation is significantly suppressed as the film thickness approaches zero (starting from relatively low bulk probabilities \cite{Jianrui2020, schmitt2024topological, Meyer2018} for large values of $h$). Notably, the maximum knotting probability lies at $h \approx 0.6 \, R_\mathrm{g,bulk}$, and is roughly independent of the chain length $N$, revealing that the models' bulk radii of gyration are the important length scales for the observed confinement-induced knotting. Furthermore, the maximum knotting probability is consistently triple the bulk value $P_\mathrm{knot}^\mathrm{max}/P_\mathrm{knot}^\mathrm{bulk} \approx 3$, exceeding the increase observed for semiflexible confined rings on a lattice, which increases by the factor $2.3$ \cite{ubertini2023spatial}. The general shape of the curves is in qualitative agreement with results observed for single-chain models in planar confinement \cite{tesi1994knot, dai2012effect, Micheletti:Macromol:2012, orlandini2013knotting, Wettermann_2023}. \newline

In contrast, entanglements are suppressed as $h \rightarrow 0$. 
As the inset to Fig. \ref{Pknotvsfilmthickness_films}(a) shows, the entanglement length $N_\mathrm{entang}$ monotonically increases with decreasing $h$, 
which is in line with observations in Ref. \cite{lee2017local}. {A likely explanation for the decrease in entanglements with increasing confinement is that individual chains pervade less volume the thinner the film confinement is \cite{brown1996entanglements}; this is further explored in the following subsection. 
We find that an analytical expression from Ref. \cite{lee2017local} accurately predicts the simulation results, which is shown in the inset to Fig. \ref{Pknotvsfilmthickness_films}(a). The expression is eq. (8) from Ref. \cite{lee2017local}: 
\begin{equation}
    N_\mathrm{entang}(h) = N_\mathrm{entang}^\mathrm{bulk} \cdot \left[ 1 + \frac{r_s}{2 \bar {h}} \cdot \ln \left( N b^2 / \bar h^2 \right) \right]^2 \, ,
\end{equation}
where $N = 1024$ is the monomer number, $b = l_b \cdot \sqrt{2.1}$ is the statistical segment length (with our model's average bond length $l_b = 0.967 \, \sigma$ and characteristic ratio $\approx 2.1$, see Ref. \cite{Meyer2018}), $\bar h$ is the effective film thickness ($\bar h = h - 1.5 \, \sigma$, where $1.5 \, \sigma$ is the necessary offset to obtain constant pressure at various thicknesses $h$), and $r_s = 0.979 \, \sigma$ is the fitted segregation length, which is approximately in line with the estimated segregation length ($0.715 \, \sigma$) based on eq. (3) of Ref. \cite{lee2017local}. \newline 

The average size of trefoil knots in terms of segments increases slightly when the confinement is on the scale of the bulk radius of gyration (Fig. \ref{Pknotvsfilmthickness_films}(b)). For each chain length $N$, the maximum trefoil knot sizes occur at slightly larger film thicknesses than the respective maximum knotting probabilities, at approximately $h \approx R_\mathrm{g,bulk}$. The relative increase of trefoil sizes due to confinement $L_\mathrm{tref}^\mathrm{max}/L_\mathrm{tref}^\mathrm{bulk} \approx 1.15$, however, is much smaller than the increase in knotting probabilities. 
{Therefore}, although knot occurrence is strongly suppressed in our thinnest simulated films, the average trefoil knot contour lengths are only weakly reduced.  
\newline

\subsection*{Connection Between Layer-Resolved Topology and Topology in Thinner Films}


Fig. \ref{Pknot_layerresolved}(a) shows the layer-resolved knotting probability alongside the squared radius of gyration and its in-plane and out-of-plane components of a relatively thick film ($h \approx 6 R_\mathrm{g,bulk}$).  Notably, all properties ($P_\mathrm{knot}$ and components of $R_\mathrm{g}^2$) align with their respective bulk values when the distance to any boundary is approximately larger than the bulk radius of gyration $y_\mathrm{CoM} \gtrsim R_\mathrm{g,bulk}$. This again reveals the bulk radius of gyration as the important length scale for confinement effects. Chains closer to the confining boundary $y_\mathrm{CoM} \lesssim R_\mathrm{g,bulk}$ are significantly more likely to be knotted. 
A slight decrease of the knotting probability for chains very close to the boundaries $y_\mathrm{CoM} \ll R_\mathrm{g,bulk}$ is also observed, similar to the decreasing knotting probability in films with very small overall thickness $h \ll R_\mathrm{g,bulk}$. The in-plane component of the squared radius of gyration ($\propto \langle R_\mathrm{g,x}^2 + R_\mathrm{g,z}^2 \rangle$) increases monotonically with increasing proximity to a boundary, while the out-of-plane component ($\propto \langle R_\mathrm{g,y}^2 \rangle$) monotonically and rapidly decreases with increasing proximity to the boundaries. This is expected, since chains with close proximity to a boundary necessarily need a significant number of monomers to be located close to the boundary, which corresponds to typically 'flattened' conformations. The overall squared radius of gyration shows non-monotonic behavior which qualitatively {coincides} 
with the knotting probability: The overall squared radius of gyration exhibits a minimum in the vicinity of the maximum knotting probability. More compact polymers are known to form knots more commonly \cite{Kantor}. However, the increase in knotting probability is {substantial} 
and does not coincide exactly with the minimum of the overall radius of gyration. This suggests that the flattened nature of polymer configurations in this regime uniquely promote knot formation. \newline 

\begin{figure}[h!]
    \centering
    \includegraphics[width=\textwidth]{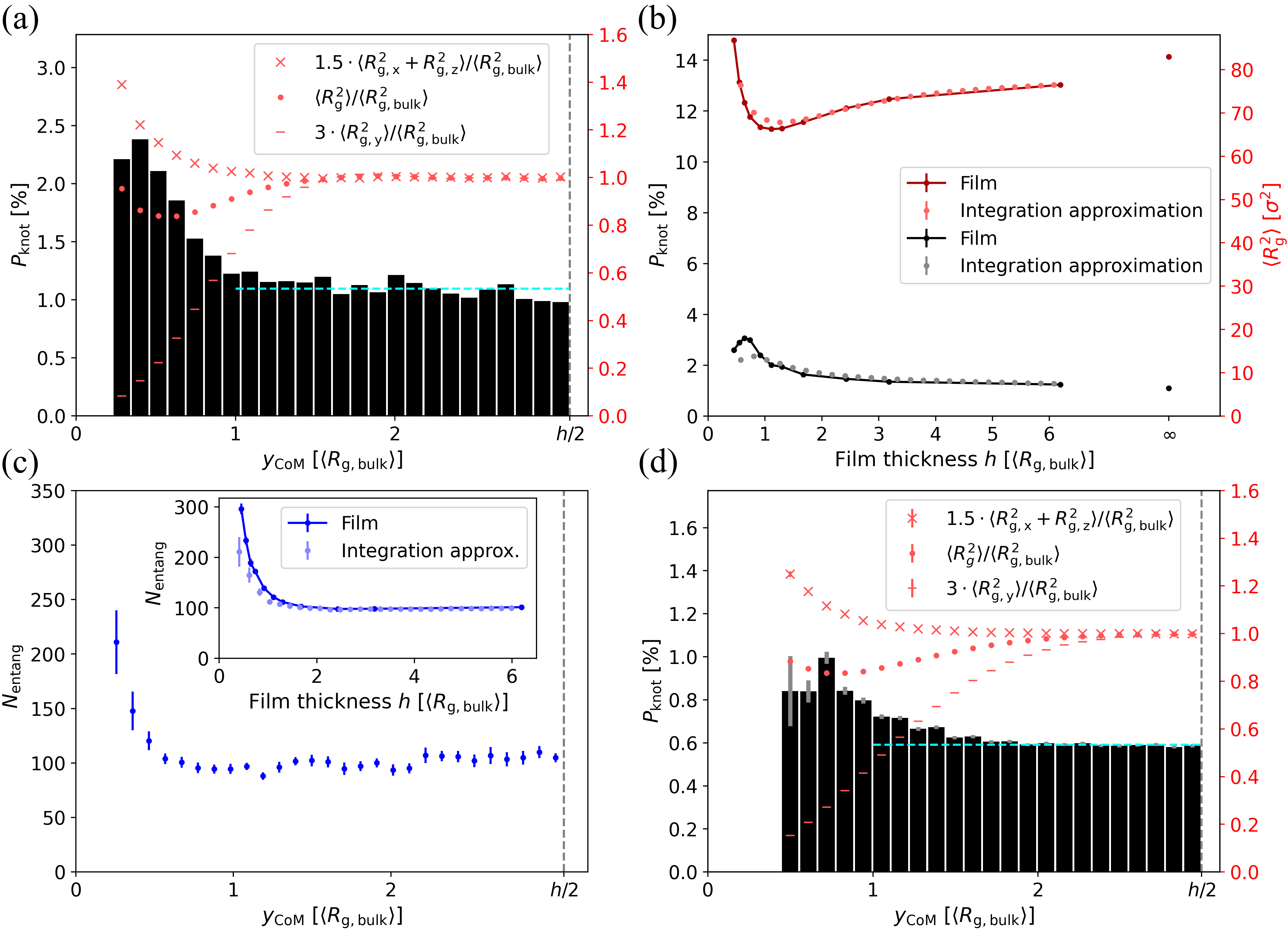}
    \caption{\textbf{Layer-resolved knotting probabilities, squared radius of gyration components and entanglement lengths, along with results of an integration scheme approximating film averages from these layer-resolved measurements.} The simulation parameters are $N=256$ number of monomers and $h=27.75 \, \sigma$ confinement thickness. Films (a-c) have a density of $\rho_\mathrm{mon} = 0.68 \, \sigma^{-3}$. For clarity and symmetry reasons, data concerning the upper half of the confinement $y > h/2$ is mirrored onto the lower half $y < h/2$ and averaged. (a) The knotting probability of the chains with central center-of-mass $y$-positions $y_\mathrm{CoM} \approx h/2$ is in line with the unconfined, bulk polymer melt value (indicated by the blue dashed line). The knotting probability increases for chains which are closer to the confinement edges, while their configurations become more flattened as seen by the radius of gyration components. (b) The knotting probability and squared radius of gyration as functions of thickness $h$ are reasonably recreated by an integration approximation, which is done by computing a weighted {average} of the layer-resolved properties from the outer edge. (c) Layer-resolved entanglement length and integration scheme results. (d) A corresponding analysis for a confined single $\theta$-chain simulation. The results are qualitatively in line with the film results in (a), although {confinement effects are} 
    weaker for single chains.}
    \label{Pknot_layerresolved}
\end{figure}

Knotting probabilities, $\langle R_g^2\rangle$ (Fig. \ref{Pknot_layerresolved}(b)) and entanglement lengths (Fig. \ref{Pknot_layerresolved}(c)) of thin films can be approximated from a thick film by means of an integration scheme. Essentially, in order to {approximate}, e.g., the {overall} knotting probability of a film with thickness $h$ {from a single simulation of a thicker film with thickness $h_0 > h$}, one {computes the} 
weighted average of the spatially resolved knotting probability of {the} thicker film {from $y = 0$ to $y = h/2$ and from $y = h_0$ to $y = h_0 - h$ (where both parts are expected to be equivalent due to symmetry).} 
Both the knotting probability as well as the squared radius of gyration are very well approximated by {this} 
integration scheme for $h \gtrsim R_\mathrm{g,bulk}$. The deviation for thinner films $h \lesssim R_\mathrm{g,bulk}$ is expected: Chains in very thin films can be influenced by both confining boundaries simultaneously, which cannot be accounted for in such an integration scheme based on a much thicker film. Despite this, the observed deviations still remain relatively small. Therefore, this integration scheme is shown to be a powerful predictor of measurable quantities in thinner films without requiring separate simulations. \newline  

The entanglement length is an additional layer-resolved quantity that shows strong similarities to its averaged counterpart as a function of thickness, as well as another example for the integration scheme. Fig. \ref{Pknot_layerresolved}(c) shows that the entanglement monotonically increases as the proximity to the wall increases beyond $y_\mathrm{CoM} \lesssim R_\mathrm{g,bulk}$. This is qualitatively in line with the increase in entanglement length with increasing confinement in the inset to Fig. \ref{Pknotvsfilmthickness_films}(a), since chains with smaller $y_\mathrm{CoM}$ tend to exhibit flattened conformations. {The volume that is pervaded by the chains is much lower near the confining boundaries as the out-of-plane component of the radius of gyration approaches zero $\langle R_\mathrm{g,y} ^2 \rangle \rightarrow 0$. This likely explains the drastic and monotonic decrease in entanglements near the boundaries, since less pervaded volume per chain means less likelihood for chains to overlap in $3d$ space \cite{brown1996entanglements}. Furthermore, our integration scheme works similarly well for the entanglement length as for the squared radius of gyration and knotting probabilities discussed above, which can be seen in the inset to Fig. \ref{Pknot_layerresolved}(c).} \newline 

In Ref. \cite{schmitt2024topological} we have shown that {single} $\theta$-chains have similar structural and topological properties as chains in the bulk, which both deviate from corresponding random walks. Here, we test this analogy for thin films:
Fig. \ref{Pknot_layerresolved}(d) shows the layer-resolved knotting probability and squared radius of gyration components {of a single $\theta$-chain confined to an equivalent slit}, analogous to Fig. \ref{Pknot_layerresolved}(a). Similar to the film, the knotting probability exhibits a maximum of approximately twice the knotting probability without confinement. The squared radius of gyration components are also qualitatively in line with the film results. However, both the knotting probability and squared radius of gyration components deviate from bulk behavior at distances slightly further away from the boundaries, compared to the film, at approximately $y_\mathrm{CoM} \lesssim 2 R_\mathrm{g,bulk}$. {This shift is likely explained by the known depletion layer of non-attractive walls \cite{fleer1993polymers}: Polymer positions close to the confining boundaries are suppressed in single-chain systems since they offer less access to conformations. However, in film systems, the density and pressure causes chains to necessarily be pushed into regions close to boundaries. Therefore, the interaction with neighboring chains {essentially} screens entropic boundary repulsion acting close to the walls. This explains the shift of deviations from bulk behavior to larger distances in single-chain system.}
{Similarly,} 
it can be seen that chains in our film {simulations} 
reach positions much closer to the wall than the single-chain {simulations} 
ever reach (i.e., there are no data points at $y_\mathrm{CoM} < \frac{1}{2} R_\mathrm{g,bulk}$ in the single-chain system, while the film still shows filled bins down to $y_\mathrm{CoM} \approx \frac{1}{4} R_\mathrm{g,bulk}$).

\subsection*{Semiflexible Polymer Films}
Introducing semiflexibility into polymer models increases their knotting \cite{schmitt2024topological} and can shift the maximum of the knotting probability as a function of film thickness. Fig. \ref{Pknotvsfilmthickness_stiffnesses} shows the knotting probability as a function of film thickness for various models ranging from flexible $B=0$ up to the semiflexible regime $B=2$. We only observe a clear maximum knotting probability in the flexible case $B=0$, while the knotting probability for $B \geq 1$ seemingly increases monotonically with increasing confinement $h \rightarrow 0$ within our simulated range. {Intuitively, one might expect stiffer polymers to approach the $2d$-limit $P_\mathrm{knot} \rightarrow 0$ faster than flexible chains, since any $180^\circ$ turns with out-of-plane components are suppressed by significant energy increases due to the bending potential. This would favor extended conformations with suppressed intra-chain crossings and thus suppressed knotting \cite{dai:Macromol:2018, virnau_2010}. However, flexible chains commonly form small clusters which cannot be penetrated by another part of the chain \cite{schmitt2024topological}, which is likely the reason for the overall increased knotting probability of non-flexible chains in bulk. This effect might be enhanced in thin films, since small clusters consisting of only few monomers could be sufficient to turn a part of the thin film completely impenetrable, thus lowering the knotting probability at very small thicknesses. Conversely, this might explain why we do not yet observe a decrease in knotting probability in strong confinement $h \rightarrow 0$ within our simulations of non-flexible chains $B \geq 1$.}

\begin{figure}[h!]
    \centering
    \includegraphics[width=.5\textwidth]{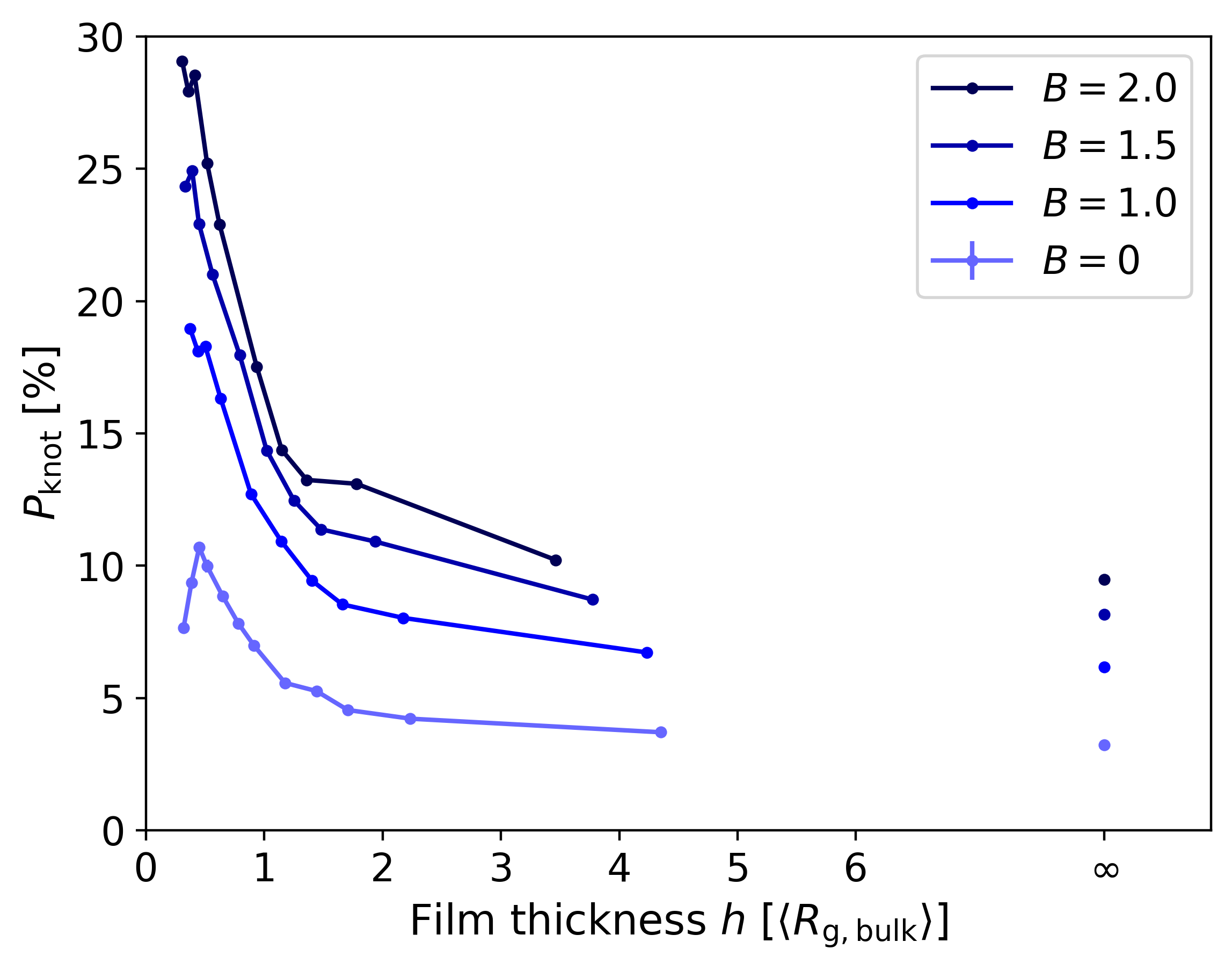}
    \caption{\textbf{Knotting probability $P_\mathrm{knot}$ 
    of confined films as a function of the film thickness $h$ at various chain stiffnesses $B$.} 
    All points correspond to a monomer number of $N=512$ and monomer density of $\rho = 0.68 \, \sigma^{-3}$. The knotting probability does not exhibit clear maxima in the implemented thicknesses when stiffness $B \neq 0$ is introduced.}
    \label{Pknotvsfilmthickness_stiffnesses}
\end{figure}

\section*{Conclusions}
We have investigated how confinement modifies the topological and conformational properties of polymer films. A central finding is that knotting probability as well as trefoil length are maximized when the film thickness is comparable to the bulk radius of gyration, $h \approx R_\mathrm{g,bulk}$. They approach bulk-like behavior in the thick film limit $h \rightarrow \infty$ and the knotting probability vanishes in the two-dimensional limit where knots cannot exist. The bulk radius of gyration consistently emerges as a critical length scale: Both global thickness-dependent trends and layer-resolved analyses of a thick film show that unique behaviors occur within distances of order $R_\mathrm{g,bulk}$. In particular, the radius of gyration is suppressed in these regions as flattened conformations occur, and this unique compactification likely causes the substantially enhanced knotting. Entanglements show a qualitatively different response. The average number of entanglements per chain decreases monotonically with confinement \cite{lee2017local}, without the non-monotonic behavior seen for knots. This distinction emphasizes that knotting is enhanced by chain compaction near walls, while entanglements are progressively suppressed as the volume occupied by individual polymers shrinks. To connect global and local perspectives, we introduce an integration scheme that reconstructs thickness-dependent properties from layer-resolved data of a single thick film. This approach captures the observed dependence of the radii of gyration, knotting probabilities and entanglement lengths on $h$, and thus provides a practical methodological framework for predicting film behavior without simulating every thickness separately. Comparison with the behavior of single chains in $\theta$-conditions show a qualitatively similar behavior, but also reveal limitations of this approach. \newline


Finally, introducing stiffness eliminates the knotting maximum within the studied thickness range. {We interpret this not as a true absence of the maximum, but as a shift to smaller thicknesses. Flexible chains form compact clusters that could block parts of the narrow films and suppress crossings, thereby reducing knotting in very thin films. Semiflexible chains are less likely to form such clusters, leaving more pathways for segments to pass around one another and form knots, leading to the strictly increasing knotting probability with increasing confinement within our simulated range of film thickness.} This interpretation highlights the sensitivity of topological effects to flexibility and suggests new directions for exploring semiflexible chains in ultrathin geometries. \newline

Our results highlight that topological and structural properties of thin films can be understood as an interplay of local properties near the confining walls and properties in the bulk.  
They clarify the contrasting responses of knots and entanglements, and provide both conceptual and methodological tools for predicting polymer topology in films.

\section*{Acknowledgements}
The authors gratefully acknowledge the computing time granted on the supercomputers MOGON II and III at Johannes Gutenberg-University Mainz as part of NHR South-West {and insightful discussions with Kostas Ch. Daoulas.} M.P.S. is grateful to the Deutsche Forschungsgemeinschaft (DFG) for funding (SFB TRR 146, Project No. 233630050). {H.M. acknowledges a generous grant of CPU time on the HPC cluster CAIUS of the mesocenter of the University of Strasbourg as well as discussions with A. Johner and J. Wittmer.}

\clearpage
\newpage

\printbibliography

\clearpage

\end{document}